\documentclass[twocolumn,showpacs,preprintnumbers,amsmath,amssymb]{revtex4}
\usepackage{graphicx}% Include figure files
\usepackage{dcolumn}% Align table columns on decimal point
\usepackage{bm}% bold math

\begin{document}

\preprint{}

\title{Electrical injection and detection of spin-polarized electrons in silicon \\through an Fe$_\text{3}$Si/Si Schottky tunnel barrier}% Force line breaks with \\

\author{Y. Ando,$^{1}$ K. Hamaya,$^{1,2}$\footnote{E-mail: hamaya@ed.kyushu-u.ac.jp} K. Kasahara,$^{1}$ Y. Kishi,$^{1}$ K. Ueda,$^{1}$ K. Sawano,$^{3}$ T. Sadoh$^{1}$ and M. Miyao$^{1}$\footnote{E-mail: miyao@ed.kyushu-u.ac.jp}}

\affiliation{$^{1}$Department of Electronics, Kyushu University, 744 Motooka, Fukuoka 819-0395, Japan}%
\affiliation{$^{2}$PRESTO, Japan Science and Technology Agency, 4-1-8 Honcho, Kawaguchi 332-0012, Japan}%
\affiliation{$^{3}$Research Center for Silicon Nano-Science, Advanced Research Laboratories, Musashi Institute of Technology, 8-15-1 Todoroki, Tokyo 158-0082, Japan.}
%\affiliation{$^{4}$Inamori Frontier Research Center, Kyushu University, 744 Motooka, Fukuoka 819-0395, Japan.}%

%\affiliation{
%%Second institution and/or address\\

\date{\today}% It is always \today, today,
\begin{abstract}
We demonstrate electrical injection and detection of spin-polarized electrons in silicon (Si) using epitaxially grown Fe$_\text{3}$Si/Si Schottky-tunnel-barrier contacts. By an insertion of a $\delta$-doped $n^{+}$-Si layer ($\sim$ 10$^{19}$ cm$^{-3}$) near the interface between a ferromagnetic Fe$_\text{3}$Si contact and a Si channel ($\sim$ 10$^{15}$ cm$^{-3}$), we achieve a marked enhancement in the tunnel conductance for reverse-bias characteristics of the Fe$_\text{3}$Si/Si Schottky diodes. Using laterally fabricated four-probe geometries with the modified Fe$_\text{3}$Si/Si contacts, we detect nonlocal output signals which originate from the spin accumulation in a Si channel at low temperatures. 
\end{abstract}
%\pacs{72.25.Dc, 72.25.Hg, 72.25.Mk}% PACS, the Physics and Astronomy
\maketitle
%\section{INTRODUCTION}
To solve critical issues caused by the scaling limit of complementary metal-oxide-semiconductor (CMOS) technologies, spin-based electronics (spintronics) has been studied.\cite{Wolf} For semiconductor spintronic applications, an electrical spin injection from a ferromagnet (FM) into a semiconductor (SC) and its detection are crucial techniques. Recently, methods for spin injection and/or detection in silicon (Si) were explored intensely\cite{Min,Igor1,Appelbaum,Jonker1,Jonker2,Grenet} because Si has a long spin relaxation time and is compatible with the current industrial semiconductor technologies. Although electrical detections of spin transport in Si conduction channels were demonstrated by two research groups,\cite{Appelbaum,Jonker2} an insulating Al$_\text{2}$O$_\text{3}$ tunnel barrier  between FM and Si was utilized for efficient spin injection and/or detection. To realize gate-tunable spin devices, e.g., spin metal-oxide-semiconductor field effect transistors (spin MOSFET),\cite{Sugahara} demonstrations of electrical spin injection and detection in Si conduction channels using Schottky tunnel-barrier contacts will become considerably important.\cite{Igor2,Ron} 

By low-temperature molecular beam epitaxy (LTMBE), we recently demonstrated highly epitaxial growth of a binary Heusler alloy Fe$_{3}$Si on Si and obtained an atomically abrupt heterointerface.\cite{Hamaya} In this letter, inserting a heavily doped $n$$^{+}$-Si layer near the abrupt interface between Fe$_\text{3}$Si and $n$-Si, we achieve an effective Shottky tunnel barrier for spin injection into Si. Using nonlocal signal measurements, we demonstrate electrical injection and detection of spin-polarized electrons in Si conduction channels though the Schottky-tunnel-barrier contacts.

The $n$$^{+}$-Si layer was formed on $n$-Si(111) ($n \sim$ 4.5 $\times$ 10$^{15}$ cm$^{-3}$) by a combination of the Si solid-phase epitaxy with an Sb $\delta$-doping process,\cite{Sugii} where the carrier concentration of the $n$$^{+}$-Si layer was $\sim$ 2.3 $\times$ 10$^{19}$ cm$^{-3}$, determined by Hall effect measurements, and $\sim$ 10-nm-thick non-doped Si layer was grown on the Sb $\delta$-doped layer. Ferromagnetic Fe$_{3}$Si layers with a thickness of $\sim$ 50 nm were grown by LTMBE at 130 $^{\circ}$C, as shown in our previous work.\cite{Hamaya} The interface between Fe$_{3}$Si and $n$$^{+}$-Si  was comparable to that shown in Ref. 11. To evaluate electrical properties of the Fe$_{3}$Si/Si Schottky contacts, we firstly fabricated two different Schottky diodes ($\sim$ 1 mm in diameter) with and without the $n$$^{+}$-Si layer between Fe$_{3}$Si and $n$-Si. Here, we define these Schottky diodes as Diode A (with the $n$$^{+}$-Si layer) and Diode B (without the $n$$^{+}$-Si layer). A schematic illustration of Diode A is shown in the inset of Fig. 1(a).

The main panel of Fig. 1(a) shows absolute values of the current density, $|I|$, as a function of bias voltage ($V_\text{bias}$) for both Diode A and B. These characteristics were reproduced for ten devices. A typical rectifying behavior of a conventional Schottky diode is seen for Diode B (blue), while we find almost symmetric behavior with respect to $V_\text{bias}$ polarity for Diode A (red). For spin injection from Fe$_{3}$Si directly into Si, the $I-V_\text{bias}$ characteristic in the reverse bias regime ($V_\text{bias} <$ 0) is important.\cite{Min,Albrecht,Ron2} We note that the reverse-bias $|I|$ ($|I_\text{rev.}|$) for Diode A is extremely large more than three orders of magnitude compared to that for Diode B. For both diodes, the temperature-dependent $|I_\text{rev.}|$ measured at $V_\text{bias} =$ - 0.3 V is representatively shown in Fig. 1(b) in order to discuss transport mechanisms in $V_\text{bias} <$ 0. For Diode B (blue), $|I_\text{rev.}|$ ($V_\text{bias} =$ - 0.3 V) is significantly reduced (more than six orders of magnitude) with decreasing temperature. By an analysis of the Arrehenius plot shown in the inset, the thermionic emission electron transport over a Schottky barrier is indicated, in which the Schottky barrier height is estimated to be $\Phi_\text{B} =$ 0.61 eV, consistent with our previous study.\cite{Hamaya} Using such contacts (without the $n$$^{+}$-Si layer), we can not detect spin-polarized carriers electrically due to additional resistance originating from a wide width of the depletion region ($\sim$ 500 nm).\cite{Min,Ron2} In contrast, we identify that $|I_\text{rev.}|$ for Diode A (red) shows almost no variation with decreasing temperature, as shown in Fig. 1(b), indicating that thermionic emission does not dominate. For Fe/AlGaAs heterostructures, Hanbicki {\it et al}. already achieved Schottky tunnel barriers with a highly-doped interface for an efficient spin injection into GaAs.\cite{Hanbicki1} The Fe/AlGaAs Schottky tunnel barriers have a very weak insulator-like temperature dependence of the conductance, indicating that the tunnel probability across the barrier is dominated by the Fermi-Dirac distribution of the conduction electron's energy in the Fe and AlGaAs.\cite{Hanbicki2} In a preliminary study, we also observed the similar weak insulator-like features of the conductance for Fe$_{3}$Si/Si Schottky barriers with a highly-doped ($\sim$10$^{18}$ cm$^{-3}$) interface. On the other hand, for the present Fe$_{3}$Si/Si Schottky barriers with a $\delta$-doped $n$$^{+}$-Si layer ($\sim$10$^{19}$ cm$^{-3}$), we could not see the weak insulator-like temperature dependence of the conductance. This may mean that the tunnel barrier becomes very thin and the wave functions of the conduction electrons between the Fe$_{3}$Si and Si are nearly overlapped. As a consequence, there is almost no influence of temperature on the tunnel probability. Thus, under the reverse-bias conditions, we can expect to achieve spin injection via a Schottky tunnel barrier consisting of the epitaxially grown Fe$_{3}$Si/Si interface, as schematically shown in the left inset of Fig. 1(a).
\begin{figure}[t]
\includegraphics[width=7.0cm]{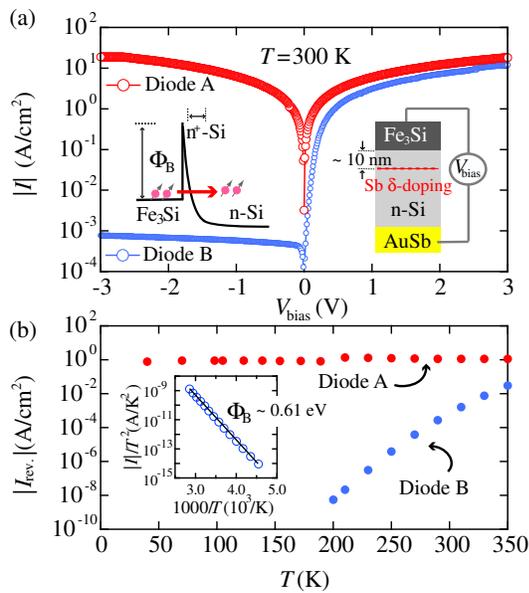}
\caption{(Color online) (a) $I-V$ characteristics of the fabricated Fe$_{3}$Si/Si(111) Schottky diodes at 300 K. The right and left insets show schematic illustrations of the Fe$_{3}$Si/Si Schottky diode structure and an energy diagram of the conduction band for Fe$_{3}$Si/Si with an $n$$^{+}$-Si layer in $V_\text{bias} <$ 0, respectively. (b) Temperature dependence of $|I|$ for the two different diodes at $V_\text{bias} =$ - 0.3 V. The inset shows an Arrhenius plot for Diode B.}
\end{figure}  
\begin{figure}[t]
\includegraphics[width=8.0cm]{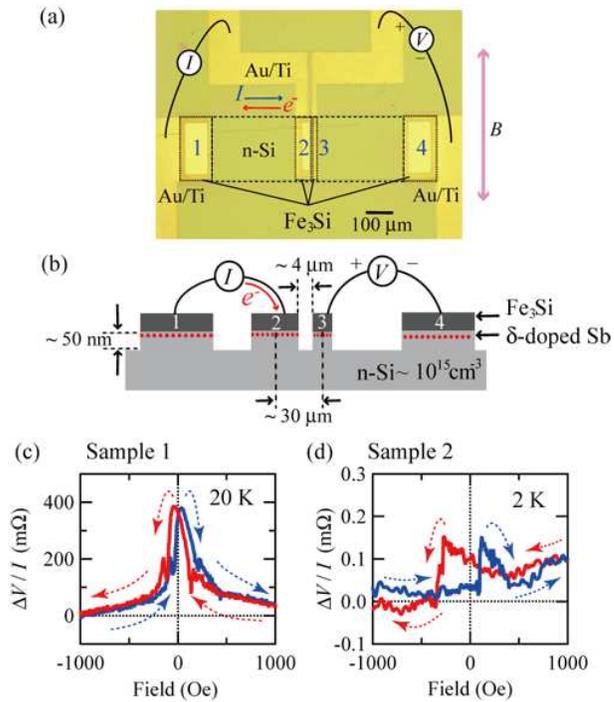}
\caption{(Color online) (a) An optical micrograph of the lateral spin device with an $n$$^{+}$-Si layer between Fe$_{3}$Si and Si for nonlocal-voltage measurements. (b) A cross-sectional diagram of the lateral device geometry. Sb $\delta$-doped layers were separated from the Si conduction channel for spin transport. Representative $\Delta$$V$/$I - B$ curves for the lateral spin device with an Fe$_{3}$Si/$n$$^{+}$-Si Schottky tunnel-barrier contact for (c) Sample 1 and (d) Sample 2, where a small linear background is subtracted from the data.}
\end{figure}  

Using the Fe$_{3}$Si/Si Schottky-tunnel-barrier contacts, we fabricated four-probe lateral spin devices, as shown in Fig. 2(a), where the lateral geometries have been utilized by lots of researchers for the detection of spin transport in semiconductor channels.\cite{Lou,Jonker2} The fabrication processes were also developed for the Fe$_{3}$Si/Si system. First, an Fe$_{3}$Si/Si mesa structure (800 $\times$ 200 $\mu$m$^{2}$) including a Si conduction channel was defined by Ar ion milling. Second, four Fe$_{3}$Si/Si contacts were defined by the ion milling, where  the Si conduction channel was situated below $\sim$ 50 nm from the Fe$_{3}$Si/Si interfaces [see-Fig. 2(b)]. Two of the Fe$_{3}$Si contacts, 2 and 3, have dimensions of 40 $\times$ 200 $\mu$m$^{2}$ and 6 $\times$ 200 $\mu$m$^{2}$, respectively [Fig. 2(a)]. The lateral distance between the contact 2 and 3 edges is $\sim$ 4 $\mu$m. To form bonding pads, a 200-nm-thick SiO$_{2}$ thin film was deposited by rf magnetron sputtering at room temperature, and then the contact holes were formed by reactive ion etching. Finally, Au/Ti pads were fabricated by electron beam evaporation and lift-off methods.

Figures 2(c) and 2(d) display representative nonlocal spin signals ($\Delta$$V$/$I$) as a function of magnetic field ($B$) for two different samples with the $n$$^{+}$-Si layers. The nonlocal voltage measurements, being reliable methods for detecting spin accumulation in a nonmagnetic channel,\cite{Lou,Jonker2,Jedema,Kimura} were performed by a conventional dc method for the current-voltage geometry shown in Fig. 2(a). External magnetic fields $B$ were applied parallel to the long axis of the Fe$_{3}$Si contacts in the film plane. Here, we define the two samples presented here as Sample 1 and Sample 2, and the lateral geometry of Sample 1 is almost identical to that of Sample 2. For Sample 1, hysteretic behavior is clearly obtained but the curve feature is somewhat different from that of the previous lateral spin valves.\cite{Lou,Jedema,Kimura} It should be noted that very large $\Delta$$V$/$I$ ($\sim$ 400 m$\Omega$) can be detected at 20 K. On the other hand, for Sample 2, we can see a spin-valve-like $\Delta$$V$/$I - B$ curve at 2 K, but $\Delta$$V$/$I$ is extremely small compared to Sample 1. For other samples without the $n$$^{+}$-Si layer, such hysteretic features could not be obtained because of the limitation of the current flow.\cite{Min,Ron2} The hysteretic nonlocal signals ($\Delta$$V$/$I$) shown in Figs. 2(c) and 2(d) indisputably exhibit the experimental demonstrations of electrical spin injection and detection in Si-based devices via a Schottky tunnel barrier. 

We now infer that the shape of the $\Delta$$V$/$I - B$ curve is affected by in-plane magnetic anisotropies for the Fe$_{3}$Si/Si(111) epilayers. For epitaxially grown Fe$_{3}$Si/Ge(111) layers, we found that there is a strong in-plane uniaxial magnetic anisotropy with a random-oriented easy axis.\cite{Ando} In addition, it is well known that such complicated in-plane anisotropies can affect in-plane magnetic configurations for epitaxially grown ferromagnetic thin films in various external magnetic fields.\cite{Ahmad,Welp} Namely, if the applied field direction, which is nearly parallel to the long axis of the wire-shaped Fe$_{3}$Si injector and detector, is deviated from its magnetic easy axis, we could not get precise parallel and anti-parallel configurations at zero field and at a magnetization switching field, respectively. These influences are probably related to the shape of the $\Delta$$V$/$I - B$ curve for Sample 1. As we can accidentally obtain a sample with the wire-shaped Fe$_{3}$Si oriented along the in-plane easy axis, we can see a spin-valve-like $\Delta$$V$/$I - B$ curve,\cite{ref.1} as shown in Fig. 2(d) (Sample 2).
\begin{figure}[t]
\includegraphics[width=8.5cm]{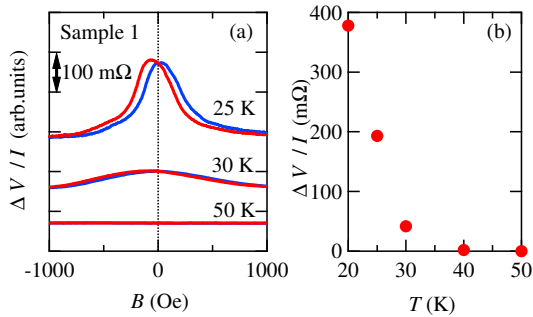}
\caption{(Color online) (a) The $\Delta$$V$/$I - B$ curves recorded at various temperatures for Sample 1. (b) $\Delta$$V$/$I$ as a function of temperature. }
\end{figure}    

Next, we focus on uneven characteristics of the $\Delta$$V$/$I$ magnitude. The $\Delta$$V$/$I$ magnitude is known to decrease with decreasing the interface resistance between ferromagnets and nonmagnets. According to a simple one-dimensional spin-diffusion model,\cite{Takahashi} the spin accumulation voltage is roughly proportional to the square of the interface resistance when the interface resistance is much smaller than the spin resistance of the nonmagnets.\cite{ref.2} We confirmed that the interface resistance for Sample 2 is quite low more than two orders of magnitude compared to that for Sample 1. Therefore, the reduction of the $\Delta$$V$/$I$ magnitude for Sample 2 is probably caused by the low interface resistance, resulting in a low efficiency of the spin injection. The precise origin of the dispersion of the interface resistance is still unclear, but the local inhomogeneity of the doped Sb ionization near the Fe$_{3}$Si/Si interface and the damage due to the device fabrication processes may influence crucially for each sample. Because of such local and slight inhomogeneous carrier concentration, we can speculate a small change in the temperature dependence of the interface resistance. This sensitive change has already been confirmed by a comparison of characteristics for diode structures ($\sim$ 1 mm in diameter) with a highly-doped ($\sim$10$^{18}$ cm$^{-3}$) interface and a $\delta$-doped $n$$^{+}$-Si ($\sim$10$^{19}$ cm$^{-3}$) interface, as described above.  

Figures 3(a) and 3(b) show temperature dependence of the $\Delta$$V$/$I - B$ curve and $\Delta$$V$/$I$, respectively, for Sample 1. The $\Delta$$V$/$I - B$ hysteresis and $\Delta$$V$/$I$ are significantly decayed with increasing temperature, and disappear completely at 50 K. We also explored the $\Delta$$V$/$I$ signal for various temperatures for Sample 2, but the hysteretic $\Delta$$V$/$I - B$ feature disappeared at 4 $\sim$ 10 K unfortunately (not shown here). If the highly efficient spin injection into Si is realized, we can see spin transport at higher temperatures because Si has a long spin-diffusion length. However, the laterally fabricated spin devices include still some problems such as the dispersion of the interface resistance, as stated in previous paragraph. We infer that the decays of the $\Delta$$V$/$I - B$ curve and $\Delta$$V$/$I$ are probably caused by a slight decrease in the interface resistance to rising temperature unfortunately, in which the decrease in the interface resistance reduces the efficiencies of the spin injection and detection. Thus, it is still insufficient that the efficiency of the spin injection and detection at the Fe$_{3}$Si/Si interface of the present devices. Further optimization of the $\delta$-doping technique, device fabrication processes, and the device geometry, is required to observe the spin transport in Si at higher temperatures.

In summary, we have demonstrated electrical injection and detection of spin-polarized electrons in silicon using epitaxially grown Fe$_\text{3}$Si/Si Schottky-tunnel-barrier contacts, where a $\delta$-doped $n^{+}$-Si layer ($\sim$ 10$^{19}$ cm$^{-3}$) is inserted near the interface between a ferromagnetic Fe$_\text{3}$Si contact and a Si channel ($\sim$ 10$^{15}$ cm$^{-3}$). By fabricating lateral four-probe geometries with the modified Fe$_\text{3}$Si/Si contacts, we detected nonlocal output signals which originate from the spin accumulation in a Si channel at low temperatures. To get a spin device with high-temperature operation, we need further optimizations of the $\delta$-doping technique, device fabrication processes, and the device geometry.  

%\noindent{{\bf Acknowledgments}}
K.H. and M.M. wish to thank Prof. Y. Shiraki, Prof. Y. Otani, Prof. T. Kimura, Prof. Y. Maeda, Prof. H. Itoh, and Prof. Y. Nozaki for their useful discussions. This work was partly supported by a Grant-in-Aid for Scientific Research on Priority Area (No.18063018) from the Ministry of Education, Culture, Sports, Science, and Technology in Japan, and PRESTO, Japan Science and Technology Agency. 

%Create the reference section using BibTeX:
%\noindent{{\bf References}}


\begin{thebibliography}{11}
\bibitem{ref.1}
The small asymmetry of the switching fields about zero-field may originate from the presence of an anti-ferromagnetic FeO thin layer at the surface of Fe$_{3}$Si.
\bibitem{ref.2}
We use Eq.(3) in Ref. 24 with the assumption that $R_{\rm Fe_{3}Si} \ll R_{\rm int.} \ll R_{\rm Si}$, where $R_{\rm int.}$ is the interface resistance between Fe$_{3}$Si and Si, and $R_{\rm Fe_{3}Si}$ and $R_{\rm Si}$ are the spin resistance for Fe$_{3}$Si and Si, respectively. Here, the low $R_{\rm Fe_{3}Si}$ is originated from the metallic conductivity and the short spin diffusion length due to the ferromagnet.
\end{thebibliography}
\end{document}